# Structured illumination in Fresnel biprism-based digital holographic microscopy

S. Hossein S. Yaghoubi[a], Samira Ebrahimi[b], Masoomeh Dashtdar[a],[*]

[a] Department of Physics, Shahid Beheshti University, Evin, Tehran 19839-69411, Iran

[b] Department of Plant and Environmental Sciences, University of Copenhagen, 1871 Frederiksberg C, Denmark

*Corresponding author: m-dashtdar@sbu.ac.ir

**Abstract**

The structured illumination (SI) architecture can be adopted in a digital holographic (DH) microscope to enhance the spatial resolution. In this paper, we propose and demonstrate a compact and simple SI method in a self-referencing common-path configuration of DH microscopy. The combination of SI pattern generated by a compact disk and common-path off-axis geometry formed by Fresnel biprism introduces a low-cost and highly stable system to image both amplitude and phase objects with a twofold improvement of the spatial resolution. Experimental results for the case of a standard test sample validate the predicted resolution enhancement compared to that of the conventional imager. The presented DH method allows for imaging living cells with strikingly improved clarity compared to the conventional DH microscopes.

## 1. Introduction

Digital holographic (DH) microscopy is a powerful label-free imaging modality which enables complete access to both amplitude and phase distribution of the sample [1,2]. DH settings can be developed and integrated with various imaging systems to simultaneously investigate the biochemical, morphological and mechanical properties of microstructures and biological cells [3-5]. In DH microscopy, the object is usually illuminated by a plane wave, and the lateral resolution of the imaging system is limited by the wavelength ($\lambda$) and the numerical aperture (NA) of the imaging system. In order to extend the spatial resolution, shorter wavelength e.g. UV and x-ray sources can be utilized which could probably damage certain types of samples such as biological tissues [6,7].

The greater NA of the imaging system is feasible by a large NA microscope objective. However, it results in the short working distance and focus depth limitation. To increase the effective NA and accordingly the lateral resolution of the microscopes, the microsphere-assisted imaging method has been proposed [8]. Structured illumination (SI) techniques are used to improve the resolving power of the imaging system by illuminating the object with a periodic pattern [9-11]. The structured light can be generated by spatial light modulators (SLMs) [12,13] and also the interference of either diffraction orders of a grating [14,15] or the deflected beams of an optical element [16] to synthesize the enlarged transfer function. So, the high-frequency information of the object is shifted within the passband of the imaging system, and multiple acquisitions are required to separate these overlapping replicas of the object Fourier transform [17-19]. These methods are time-consuming due to the requirement of recording at least two images, and any mechanical vibrations could increase the temporal noise level. Therefore, the high temporal stability of the settings is a desired feature for any SI implementation, and this key feature can be provided by the common-path geometry instead of previously proposed two-beam systems [9,20]. In the recent proposed shearing-based DH configuration, the high-resolution images are obtained by illuminating the sample using a structured pattern by a grating [21].

In this paper, we present a compact and simple SI method in a highly stable self-referencing common-path geometry of DH microscopy employing a Fresnel biprism. The proposed setting enables to enhance the resolution of DH techniques using a structured light generated by a grating. This approach requires two off-axis acquisitions of the SI and the conventional illumination (CI) holograms. To separate the high-frequency information carried by the first diffraction order of the grating from low-frequency information, only a simple subtraction is applied. The intensity reduction between the two interfered beams may be appeared in the reflection-based interferometers, leading to a reduction in the contrast of the fringes [21]. Due to the same intensity and near optical path length of two waves deflected by the biprism, high contrast and robust-to-vibration linear fringes are generated and this capability can improve the precision of the measurements. The curvature of both beams is the same in this method, which ignores the effect of quadratic phase factor in the reconstructed phase maps. Thus, the common-path setup based on Fresnel biprism is an efficient way to eliminate the uncorrelated and undesirable phase variations. The adjustable fringe modulation and also easy implementation of this quantitative DH system make it ideal for high-resolution 3D phase imaging. The proposed method can be coupled with different types of microscopes for multimodal imaging capability.

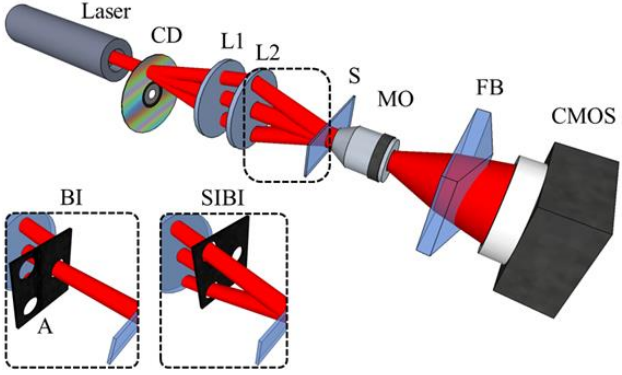

**Fig. 1.** Schematic diagram of the proposed SI technique combined with common-path biprism setting. CD: compact disc; L1 and L2: lenses; S: sample; MO: microscope objective; FB: Fresnel biprism. The dashed line box is illustrated in down subfigures and using a movable aperture (A), we can switch between the conventional illumination (BI box) and the structured illumination (SIBI box) schemes.

## 2. Experimental setup

The proposed system is a modified common-path setting based on biprism interferometer (BI) [22,23], as shown in Fig. 1. The light from a He-Ne laser of wavelength $\lambda = 632.8$ nm illuminates a clear compact disc (CD) as the grating with a period of 1.6 μm to generate the diffraction orders. Three central diffracted orders $(0,\pm1)$ are projected onto the sample (S) plane by a two-lens system (L1-L2) with the same focal length of $f = 50$ mm to generate the SI. The illuminated object is magnified by a 5× /0.1 NA microscope objective (MO). Then, the exiting beam from the MO meets the Fresnel biprism (FB) with a refringence angle of $\theta = 0.07$ radian and a refractive index of $n = 1.51$ where the wavefront splits into two interfering beams, creating the off-axis arrangement. A CMOS sensor is used to record the interference fringes. It is noted that the sample is placed in a plane conjugate to the grating. Also, L1 is located 12 mm from the grating, the separation distance between L1 and L2 is 42 mm, and the illumination angle of ±1 onto the sample is 7.6 degrees. The structured light moves the high-frequency components into the observable region. Therefore, a SI hologram contains both the high-frequency and the low-frequency information which are mixed together, and a conventional hologram is required to only extract the high-frequency component from the SI hologram. To address this desire, we have placed a multi-aperture (A) composed of two switchable apertures after L2 for blocking either the higher frequencies or the lower frequencies. By blocking the ±1 diffraction orders, the conventional hologram is recorded (see the left-down box in Fig. 1 (BI) which is the magnified section, specified by a dashed line in the original figure). Then, the aperture is moved to block only the central region, as depicted in the right-down box of Fig. 1 (SIBI), and after passing ±1 terms through the aperture, the SI hologram is captured. The integrated configuration of SI and biprism interferometer (SIBI) requires only two holograms without moving the grating, and its common-path architecture provides a highly stable system which is an important criterion for multi-shot imaging techniques.

## 3. Theory and numerical reconstruction

In conventional DH microscopy using the Fresnel biprism, an off-axis geometry is implemented in which two deflected waves are traveled with an angle that is specified by the refractive index and the refringence angle of the biprism. This configuration is suitable for investigation of the sparse samples as well as the confluent specimens by placing it in the half of the field of view. The part of the beam which is modulated with the complex information of the sample ($O'$), acts as the object beam for the other part, known as the reference beam. Two emerging beams are recombined and interfered at the sensor plane to create the off-axis hologram. The central edge diffraction effect of biprism can be avoided by proper selection of the field of view. The Fourier transform of the recorded hologram by the sensor is given by

$$\tilde{I}(\mathbf{K}) = DC(\mathbf{K}) + \tilde{O}'^{*}(\mathbf{K}) \otimes \delta(\mathbf{K} - \gamma) + \tilde{O}'(\mathbf{K}) \otimes \delta(\mathbf{K}+\gamma) \quad (1)$$

where $\otimes$ is the convolution operator, $\mathbf{K}$ vector denotes the 2D spatial frequency, and the magnitude of $\gamma$ vector is $2b(n-1)\theta/\lambda L$ dealing with the carrier spatial frequency of the hologram fringes induced by deflections of both object and reference beams. $b$ and $L$ are the MO-to-biprism and the MO-to-sensor distances, respectively. By moving the biprism along the optical axis and changing b parameter, the spatial frequency of fringes can be easily adjusted for a successful filtering procedure in the Fourier domain [22]. The DC term in Eq. (1) is located at the center of the spatial frequency coordinates. The second and third terms represent the real and virtual images of the object, which contain both the amplitude and phase information. The DC term and twin images are suppressed by Fourier analysis technique, and the image containing the complex amplitude of the sample is isolated from the hologram's spectrum. If the object is illuminated by a sinusoidal pattern, each term in the Fourier transform of Eq. (1) contains two other copies, which are symmetrically distributed around the main term and separated by the spatial frequency of the sinusoid. The periodic pattern in the object plane is denoted by $t(\xi,\eta) = 1 + m/2\cos[2\pi\eta/Mp_\eta]$, where $p_\eta$ is the period of the grating, $(\xi,\eta)$ are the lateral coordinates of the object plane, and $M$ is the magnification of the two-lens system. Also, $m$ represents the coefficient of the first diffraction order. By denoting $O(\xi,\eta)$ as the transmission of the object, complex amplitude wavefront in the object plane is $O'(\xi,\eta) = O(\xi,\eta)t(\xi,\eta)$ which consists of three phase-shifted replicas of the object spectrum in the Fourier domain. The SI wavefront is split into two overlapping beams by the biprism, and the SI hologram of the magnified object is recorded at the sensor plane. The filtered Fourier spectrum of this SI hologram (equivalent to the second term in Eq. (1)) is expressed as

$$U(\mathbf{K}) = O'^*(\mathbf{K}) \otimes \delta(\mathbf{K}-\boldsymbol{\gamma}) + \frac{m}{4}\tilde{O}'^*(\mathbf{K}) \otimes \delta(\mathbf{K}-\boldsymbol{\gamma}-\boldsymbol{\alpha}) \\ + \frac{m}{4}\tilde{O}'^*(\mathbf{K}) \otimes \delta(\mathbf{K}-\boldsymbol{\gamma}+\boldsymbol{\alpha}), \quad (2)$$

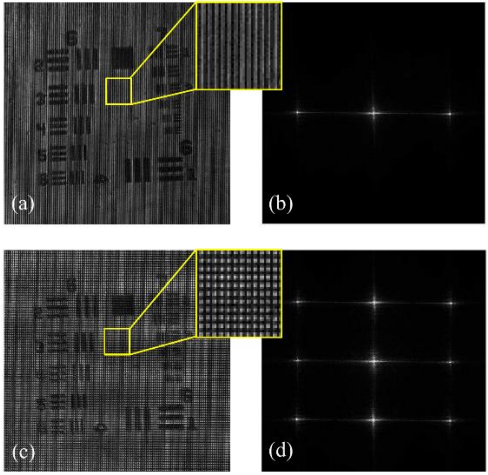

**Fig. 2.** Experimental results obtained by recording groups 6 and 7 of a 1951 USAF resolution test target. (a) The captured hologram by BI system and (b) its corresponding Fourier spectra. (c) The hologram obtained by the SIBI system and (d) its corresponding Fourier spectra. The zoomed-in areas in (a) and (c) demonstrate the CI and SI interference fringes, respectively.

Where $\boldsymbol{\alpha}$ vector magnitude is denoted by $1/M_t p_\eta$, and $M_t$ is the total lateral magnification between the grating and image plane ($M_t = M \times M_O$). The magnification factor between the object and the image plane is denoted by $M_O$. The theoretical lateral resolution improvement of the reconstructed SI images is determined by the illumination angle between diffraction orders of the grating ($\theta_{illum}$). The achievable lateral resolution by the SI system is $\delta = \kappa\lambda/[NA + \sin(\theta_{illum})]$ [24], where the constant parameter $\kappa$ is determined by the experimental parameters [25]. When $\sin(\theta_{illum}) = NA$, the maximum resolution enhancement can be obtained.

The angular spectrum propagation (ASP) method and a numerical autofocusing technique are employed to retrieve the complex amplitude of the recorded holograms [2, 26]. After recording two images including one hologram for CI and the other for SI pattern, digitally filtered and propagated angular spectra of the holograms are combined to obtain the high-resolution image. Firstly, DC term and twin image are suppressed in the spectra of both holograms. The cropped low-frequency information of the object from the CI hologram is then shifted to the center of the Fourier plane to compensate $\boldsymbol{\gamma}$. To guarantee no overlap between the two high-frequency terms and the low-frequency information of the object in the spectrum of SI hologram, the spectrum of the CI hologram is subtracted. Then, the high-frequency components in Eq. (2), second and third terms, are isolated by peak-detection and set back to their respective positions in the Fourier domain by properly shifting. By mixing the high-frequency information with the low-frequency information, the high-resolution object spectrum can be obtained. Once the high-resolution object spectrum is determined, the ASP approach is applied to retrieve the object complex amplitude at the image plane, $A(x,y)$, and consequently phase information of the object, $\varphi(x,y) = \tan^{-1}[\text{Im}(A)/\text{Re}(A)]$. The continuous phase distribution of the object is acquired using a fast phase unwrapping algorithm [27].

### 4. Experimental results

Our proposed method has been initially verified for enhanced resolution imaging utilizing an amplitude 1951 USAF resolution test target, as a calibration object. The comparison between the CI and extended resolution amplitude images is carried out. Figures 2(a) and 2(c) illustrates the captured holograms of the test chart which is illuminated by uniform light and SI pattern. The recorded fringes of the CI-based BI method are magnified in the yellow inset of Fig. 2(a). Figure 2(b) depicts the associated angular spectrum of the CI recorded hologram. The horizontal fringes created by the grating are overlapped by off-axis illumination fringes (Fig. 2(c)) and lead to Fourier spectra of the SI hologram in Fig. 2(d). The object spectrum in Fig. 2(b) is filtered out and then relocated to the center of the Fourier domain (Fig. 3(a)). The high-resolution spectrum of Fig. 3(c) is the synthesized aperture, generated by subtracting the conventional spectrum from the SI spectrum and properly mixing the filtered spectrum of Fig. 2(b) and 2(d)-2(b). Then the complex amplitudes of the reconstructed image for CI (Fig. 3(b)) and SI (Fig. 3(d)) can be compared. To compare the resolution limit of each method, the finest element that can still be resolved is determined. According to the Rayleigh criterion, the details can be resolved when the minimal dip is at least 26.3% of the maxima. The smallest resolvable element with the conventional imager in Fig. 3(b) is element 3 in group 6 (80.6 lp/mm). However, the finest details that can be resolved by the SI imager (Fig. 3(d)) is element 3 in group 7 (161.3 lp/mm). Therefore, the presented setting achieves a twofold improvement of the spatial resolution. One-dimensional profiles along the vertical direction through the horizontal bars of elements 6-3 and 7-3 are represented in Fig. 3(e) and 3(f), respectively. Obviously, the horizontal bars of the 7-3 element cannot be resolved by CI. Additionally, it should be noticed that the vertical bars of all frequencies in group 7 are not resolvable by both CI and SI measurements with resolution enhancement in the vertical orientation.

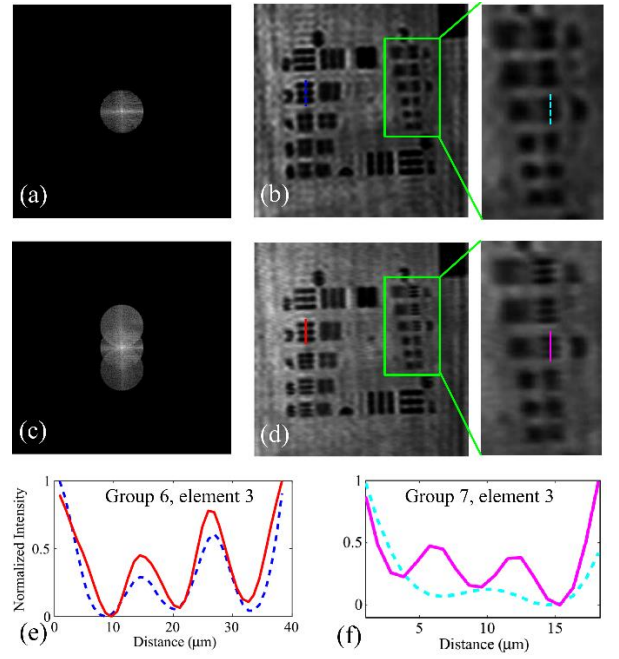

**Fig. 3.** The reconstructed amplitude images of the USAF test target which show group 6 and 7. (a) The spectrum of CI hologram and (b) the retrieved amplitude distribution with the zoomed-in portion of group 7 specified by a green rectangle. (c) The high-resolution spectrum which is produced by combining (a) with the high spatial frequencies of the SI pattern and its reconstructed amplitude image (d). The cross-section comparison between BI and SIBI for element 6-3 (e) and element 7-3 (f) along colored dashed lines in (b) and colored lines in (d).

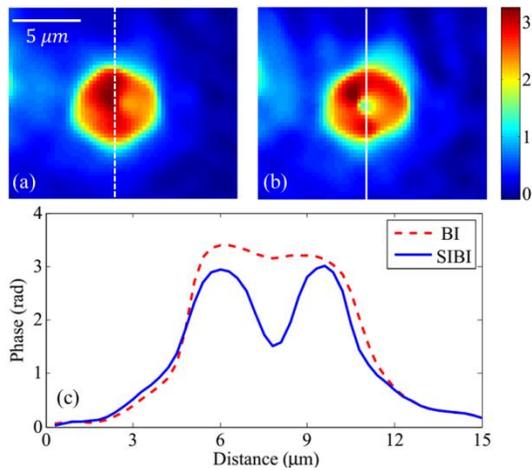

**Fig. 4.** 2D representations of reconstructed phase in radian for conventional (a) and structured (b) illumination of an RBC. (c) Cross-sections of the phase profiles of an RBC corresponding with red dashed line and blue line for BI and SIBI systems, respectively.

Another experiment has been carried out to show the resolution enhancement for imaging the biological cells by the proposed method. For this demonstration, a thin smear of healthy red blood cells (RBCs) is placed in the object plane, then CI and SI holograms are captured and the phase distributions are retrieved. The reconstructed phase map of conventional BI and super-resolved SIBI images are illustrated in Fig. 4(a) and 4(b), respectively. The cross-sections of the cell in Fig. 4(c) represent the comparison between BI (red dashed line) and SIBI (blue line) images. The result completely confirms the expected resolution enhancement of SIBI compared to that of the conventional BI system, since the biconcave shape of the cell is clearly detected by the SIBI technique.

To measure the stability of the system, a glass slide is located in the object plane, and sequential holograms are recorded for 60s at 40 Hz. Phase reconstruction is performed for a 100×100-pixel region of each frame and the path length variation is evaluated in comparison with a reference frame. Then, the standard deviation for each pixel across all frames is calculated, and the average value of the standard deviations $\sigma_\mu$ is determined for measuring the setup stability. By placing the setting on a table without isolation vibration, the holograms are captured for the CI scheme. Figure. 5 demonstrates the histogram of these standard deviation values with a mean value $\sigma_\mu = 0.62$ nm. Based on this measurement, the configuration represents a viable sub-nanometer temporal stability to detect the nanometer-scale optical path length changes of micron-sized specimens such as living cells.

## 5. Conclusion

In conclusion, an efficient architecture for integrating a SI technique using a grating and double-lens system into the conventional DH microscopy system based on common-path geometry employing a Fresnel biprism is presented. This method provides an improved resolution quantitative phase imaging by two acquisitions. The setting is highly stable, easy-to-implement and low-cost. It can easily be integrated with any optical microscope for label-free high-resolution 3D imaging. The lateral resolution prediction is evaluated using a standard test sample, and the results demonstrate a resolution improvement factor of about two of the SI system of biprism-based DH microscopy over the CI-DH system. The performance of the proposed method is evaluated by imaging the living cells which resolves the finer details compared to a CI-DH microscope. Future work lies in developing the system for single-shot imaging of dynamic phenomena.

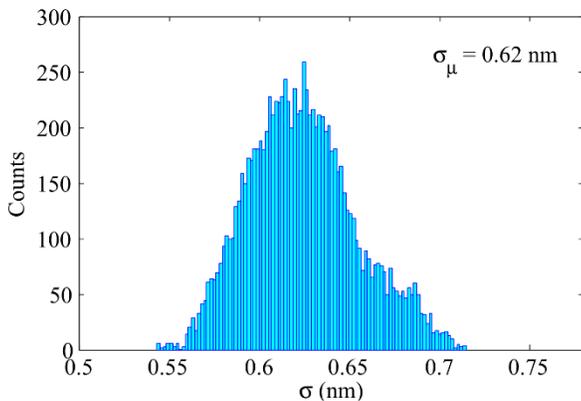

**Fig. 5.** Experimental results for the temporal stability of the BI system. Histogram of standard deviations of fluctuations of 100×100-pixel region recorded at 40 Hz for 60 s. $\sigma_\mu$ demonstrates the mean value of standard deviations of the setup stability.